\documentclass[aps,prl,twocolumn,amsmath,amssymb,superscriptaddress,nobibnotes,showpacs]{revtex4-1}

\usepackage{graphicx}
\usepackage{dcolumn}
\usepackage{bm}
\usepackage{amsmath}
\usepackage{amssymb,amsthm}
\usepackage{epsfig,color}
\usepackage[sans]{dsfont}

\definecolor{nblue}{rgb}{0.2,0.2,0.7}
\definecolor{ngreen}{rgb}{0.2,0.6,0.2}
\definecolor{nred}{rgb}{0.7,0.2,0.2}
\definecolor{nblack}{rgb}{0,0,0}

\newcommand{\tr}{\text{tr}}

\def\x{\mathrm{x}}
\def\g{\mathrm{guess}}
\def\y{\mathrm{y}}
\def\z{\mathrm{z}}

\def\s{ \vec{\sigma}}

\def\v{\mathrm{v}}

\def\P{\mathcal{P}}

\def\tr{\mbox{tr}}
\def\bea{\begin{eqnarray}}
\def\eea{\end{eqnarray}}

\begin{document}


\title{Minimum-Error Discrimination of Qubit States - Methods, Solutions, and Properties}

\author{Joonwoo Bae} \email{bae.joonwoo@gmail.com}
\affiliation{Center for Quantum Technologies, National University of Singapore, 3 Science Drive 2, Singapore 117543 and, \\
ICFO-Institut de Ci\'encies Fot\'oniques, Mediterranean Technology Park, 08860 Castelldefels (Barcelona), Spain,}

\author{Won-Young Hwang}
\affiliation{Department of Physics Education, Chonnam National University, Gwangju 500-757, Korea}
\date{\today}


\begin{abstract}
We show a geometric formulation for minimum-error discrimination of qubit states, that can be applied to arbitrary sets of qubit states given with arbitrary \emph{a priori} probabilities. In particular, when qubit states are given with equal \emph{a priori} probabilities, we provide a systematic way of finding optimal discrimination and the complete solution in a closed form. This generally gives a bound to cases when prior probabilities are unequal. Then, it is shown that the guessing probability does not depend on detailed relations among given states, such as angles between them, but on a property that can be assigned by the set of given states itself. This also shows how a set of quantum states can be modified such that the guessing probability remains the same. Optimal measurements are also characterized accordingly, and a general method of finding them is provided. 



\end{abstract}
\pacs{03.65.Ud, 03.67.Hk, 03.65.Wj}


\maketitle

Discrimination of quantum states is a fundamental processing to extract information encrypted in collected quantum states. In practical applications, its framework characterizes communication capabilities of encoding and decoding messages via quantum states \cite{ref:hel} \cite{ref:el}. The process of distinguishing quantum states is generally a building block when quantum systems are applied to information processing, in particular for communication tasks \cite{ref:hel}. Its usefulness as a theoretical tool to investigate quantum information theory has also been shown, according to the recent progress along the line, in secure communication, randomness extraction in classical-quantum correlations \cite{ref:ren}, and semi-device independent quantum information tasks \cite{ref:semi}.

Despite of much effort devoted to discrimination of various sets of quantum states so far, see reviews in Ref. \cite{ref:rev} for theoretical and experimental developments, however, apart from the general method of two-state discrimination shown in 1976 \cite{ref:hel} or restricted cases where some specific symmetry exists in given quantum states, e.g. Ref. \cite{ref:eldar}, little is known in general about optimal discrimination of quantum states, see also the review in Ref. \cite{ref:pro} about the progress. For instance, {\it the next simplest example that comes after the two-state discrimination is an arbitrary set of three qubit states, for which no analytic solution is known yet}. When arbitrary quantum states given, a general method for the optimal discrimination has been a numerical procedure e.g. Ref. \cite{ref:sdpbook} and only numerically approximate the exact solutions. Needless to speak about the importance in its own right, the lack of a general method for state discrimination even in simple instances is, due to its fundamental importance, obviously and also potentially a significant obstacle preventing further investigation in both quantum information theory and quantum foundation. 

In the present work, we provide a remarkable progress along the long-standing problem of minimum-error state discrimination, in particular, for arbitrarily given sets of qubit states. This would lead to significant improvement along the related problems in quantum information, see those in the review in Ref. \cite{ref:rev} or the recent applications of state discrimination e.g Ref. \cite{ref:semi}. We show a geometric formulation of optimal discrimination of qubit states by analyzing optimality conditions. This approach is called \emph{the complementarity problem} in the context of semidefinite programming. We provide the guessing probability, i.e. the maximal probability of making correct guesses, in a closed form for cases where arbitrary qubit states, among which no symmetry may exist, are given with equal \emph{a priori} probabilities. The geometric formulation also applies to other cases of unequal \emph{a priori} probabilities, and we characterize optimal measurements accordingly. From these results, it is shown that the guessing probability does not depend on detailed relations among given states in general but the property assigned by the set of given states. This also shows how a set of states can be modified such that the modification cannot be recognized in the discrimination task in terms of the guessing probability.

For the purpose, let us briefly summarize the minimum-error discrimination in the context of a communication scenario of two parties, Alice and Bob. They have agreed with $N$ alphabets $\{\x\}_{\x=1}^{N}$ and states $\{\rho_{\x} \}_{\x=1}^{N}$, as well as \emph{a priori} probabilities $\{ q_{\x} \}_{\x=1}^{N}$. Alice's encoding works by mapping alphabet $\x$ to state $\rho_{\x}$, and relating states with \emph{a priori} probabilities $\{q_{\x}\}_{\x=1}^{N}$. This can be seen as Alice's pressing button $\x$ with probability $q_{\x}$, and then Bob's guessing among states $\{\rho_{\x}\}_{\x=1}^{N}$ given with \emph{a priori} probabilities $\{q_{\x}\}_{\x=1}^{N}$, which we write by $\{q_{\x},\rho_{\x} \}_{\x=1}^{N}$. 

Bob's discrimination of quantum states is described by Positive-Operator-Valued-Measure (POVM) elements: $\{ M_{\x}\geq 0\}_{\x=1}^{N}$ satisfying $\sum_{\x} M_{\x} =I$ (completeness). Let $P_{B|A}(\x |\y)$ denote the probability that Bob has a detection event on $M_{\x}$ that leads to the conclusion that $\rho_{\x}$ is given, while a state $\rho_{\y}$ is provided by Alice's sending message $\y$. This is computed as follows, $P_{B|A}(\x|\y) :=P(\x|\y) = \tr[M_{\x} \rho_{\y}]$. The figure of merit is the maximal probability that Bob makes a correct guess on average, called \emph{the guessing probability},
\bea 
P_{\g } = \max_{\{ M_{\x}\}_{\x=1}^{N} } \sum_{\x} q_{\x} \tr[M_{\x} \rho_{\x} ], ~~~\sum_{\x}M_{\x}=I, \label{eq:gp}
\eea
where the maximization runs over all POVM elements. This naturally introduces the discrimination as an optimization task. 

In fact, the above can be put into the framework of semidefinite programming \cite{ref:sdpbook}. A useful property in this approach is that a given maximization (minimization) problem can be alternatively described by its dual, a minimization (maximization)  problem. The dual problem to the maximization in Eq. (\ref{eq:gp}) is obtained as follows,
\bea
P_{\g} = \min_{K} \tr[K], ~~~ K\geq q_{\x} \rho_{\x},~~\forall\x=1,\cdots,N. \label{eq:dgp}
\eea
In this case, the minimization works to find a single parameter $K$ which then gives the guessing probability, e.g. the approach in Ref. \cite{ref:dual}. 

For convenience, we call the problem in Eq. (\ref{eq:gp}) as the primal, with respect to the dual in Eq. (\ref{eq:dgp}). Note that solutions of two problems do not generally coincide with each other. The fact that, in this case, the guessing probability can be obtained from both primal and dual optimizations follows from the property called strong duality. This holds when both primal and dual problems have a non-empty set of parameters satisfying given constraints, which is referred to as the feasible problems. Once both problems are feasible, the strong duality holds, and then it follows that solutions from both problems coincide with each other. 

Apart from solving those optimization problems, there is another approach called \emph{a complementarity problem}. This collects optimality conditions that parameters of both primal and dual problems should satisfy, in order to give optimal solutions. Then, any set of parameters satisfying optimality conditions immediately provide optimal solutions of primal and dual problems. As more parameters are taken into account, the approach is not considered to be easier, however, the advantage lies at its usefulness to find general structures of a given problem. 

In the semidefinite programming formulation, the optimality conditions can be summarized by the so-called Karus-Khun-Tucker (KKT) conditions. For quantum state discrimination, they are given by, together with two constraints in Eqs. (\ref{eq:gp}) and (\ref{eq:dgp}), 
\bea 
K &=& q_{\x} \rho_{\x} + r_{\x} \sigma_{ \x},~~\mathrm{and}   \label{eq:kkt1} \\
&& r_{\x} \tr [\sigma_{\x} M_{\x}] =0 ,~~\forall ~\x=1,\cdots, N \label{eq:kkt2}
\eea
for a set of $complementary$ $states$ $\{ \sigma_{\x}\}_{\x=1}^{N}$ with non-negative coefficients $\{r_{\x}\geq 0\}_{\x=1}^{N}$ and POVM elements $\{M_{\x} \}_{\x=1}^{N}$. Once states $\{r_{\x}, \sigma_{\x}\}_{\x=1}^{N}$ and measurements $\{ M_{\x}\}_{\x=1}^{N}$ satisfying these conditions are found, they are automatically optimal to give solutions in both primal and dual problems. From the fact that the strong duality holds in this case, it is clear that the guessing probability is obtained from either of the problems. Note that the former condition in Eq. (\ref{eq:kkt1}) is called the Lagrangian stability, and shows that there exists a single operator $K$ that can be decomposed into $N$ different ways. The latter one in Eq. (\ref{eq:kkt2}) is the complementary slackness that shows the orthogonality relation between primal and dual parameters.
 
The particular usefulness of KKT conditions here is that, as it is shown in the above, they separate the guessing probability (i.e. $\tr[K]$ in Eq. (\ref{eq:dgp})) from optimal measurements: a single operator $K$ solely characterizes the guessing probability, and optimal measurements themselves are independently expressed in Eq. (\ref{eq:kkt2}). The operator $K$ can be explained to have $N$ decompositions with $q_{\x} \rho_{\x}$ and $r_{\x} \sigma_{\x}$ for $\x=1,\cdots, N$. Optimal measurements are described as POVM elements orthogonal to states $\{\sigma_{\x}\}_{\x=1}^{N}$ for each $\x$. The discrimination problem is then equivalent to finding states $\{ \sigma_{\x}\}_{\x=1}^{N}$ to fulfill these conditions. 

We now show a geometric formulation to find complementary states. Let us first define the polytope of given states $\{ q_{\x}, \rho_{\x}\}_{\x=1}^{N}$ denoted as $\P(\{ q_{\x}, \rho_{\x}\}_{\x=1}^{N})$ in the underlying state space, in which each vertex corresponds to $q_{\x} \rho_{\x}$. It is useful to rewrite the condition in Eq. (\ref{eq:kkt1}) as,
\bea 
q_{\x} \rho_{\x} - q_{\y} \rho_{\y} = r_{\y} \sigma_{\y} - r_{\x} \sigma_{\x},~~~\forall~\x,\y\label{eq:kkt1p}
\eea
This shows that two polytopes $\P(\{ q_{\x}, \rho_{\x}\}_{\x=1}^{N})$ of given states and $\P(\{ r_{\x}, \sigma_{\x}\}_{\x=1}^{N})$ of complementary states, which to search for, are actually congruent. Thus,  the structure of complementary states is already determined from given states $\{ q_{\x}, \rho_{\x}\}_{\x=1}^{N}$. Once the state geometry is clear e.g. \cite{ref:bookgeo}, the formulation can be applied.



For qubit states, their geometry can generally be described in the Bloch sphere in which the distance measure is given by the Hilbert-Schmidt norm. In what follows, we restrict our consideration to qubit states and apply the geometric formulation to discrimination among them. For a qubit state $\rho_{\x}$, we write its Bloch vector as $\mathrm{v}(\rho_{\x})$, with which $\rho_{\x} = (I+\v (\rho_{\x}) \cdot \vec{\sigma})/2$ where $\s= (X,Y,Z)$ Pauli matrices $X$, $Y$, and $Z$. 

We first characterize the general form of optimal measurements for qubit-states discrimination, from the KKT condition in Eq. (\ref{eq:kkt2}). Suppose that $r_{\x}>0$, otherwise, the measurement can be arbitrarily chosen. To fulfill the condition, it is not difficult to see that optimal POVM elements are either of rank-one \cite{ref:note1} or the null operator. If $\sigma_{\x} =  |\psi_{\x} \rangle \langle \psi_{\x}|$ then $M_{\x} = m_{\x} |\psi_{\x}^{\perp} \rangle \langle \psi_{\x}^{\perp} |$ with coefficients $m_{\x}$, where it holds that $\v(\psi_{\x}) =- \v(\psi_{\x}^{\perp})$. If a state $\sigma_{\x}$ is not of rank-one, the only possibility to fulfill the KKT condition in Eq. (\ref{eq:kkt2}) is that the measurement corresponds to the null operator, i.e. $M_{\x}=0$. In fact, optimal discrimination sometimes consists of a strategy that makes a guess without actual measurement  \cite{ref:hunter}. Note that, however, as measurements are done in most cases (otherwise, the completeness of POVM elements in the following is not fulfilled), one does not have to immediately assume that $\{\sigma_{\x}\}_{\x=1}^{N}$ are not of rank-one from the beginning. Then, for cases where measurements are done, corresponding complementary states must be of rank-one - otherwise, the orthogonality in Eq. (\ref{eq:kkt2}) cannot be fulfilled. 

Once states $\{\sigma_{\x}\}_{\x=1}^{N}$ are found, optimal measurements are automatically obtained. What remains is that POVM elements fulfill the completeness, $\sum_{\x}M_{\x}=I$, or equivalently in terms of Bloch vectors, $\sum_{\x} m_{\x} v(\psi_{\x}^{\perp})=0$ while $\sum_{x} {m_{\x}}=2$ with $m_{\x}\geq0$, $\forall x$. This refers to finding a convex combination $\{m_{\x}\}_{\x=1}^{N}$ of Bloch vectors $\{ \v(\psi_{\x}^{\perp}) \}_{\x=1}^{N}$ such that it results to zero, i.e. the origin of the Bloch sphere. This is equivalent to the condition that the convex hull of Bloch vectors $\{ \v(\psi_{\x})  \}_{\x=1}^{N}$ of complementary states contain the origin of Bloch sphere. As we will show later, this is always fulfilled by complementary states. To summarize, once complementary states are found, it is automatic to have optimal POVMs as their Bloch vectors are determined and the completeness is also straightforward.  

  


We can thus proceed to construction of complementary states in the Bloch sphere for qubit states $\{q_{\x},\rho_{\x} \}_{\x=1}^{N}$. Let us identify the polytope $\P(\{ q_{\x}, \rho_{\x}\}_{\x=1}^{N})$ in the state space as the convex hull of their Bloch vectors $\{q_{\x},\v( {\rho_{\x}}) \}_{\x=1}^{N}$, so that each vertex $q_{\x}\rho_{\x}$ corresponds to the Bloch vector $q_{\x}\v( {\rho_{\x}})$.  Then, the task is to find the polytope $\P(\{ r_{\x}, \sigma_{\x}\}_{\x=1}^{N})$ of complementary states that is congruent to  $\P(\{ q_{\x}, \rho_{\x}\}_{\x=1}^{N})$ in the Bloch sphere. Moreover, as it is shown, most of complementary states $\{\sigma_{\x}\}_{\x=1}^{N}$ are of rank-one and themselves lie at the border, thus, the polytope $\P(\{\sigma_{\x} \}_{\x=1}^{N})$ only of them is maximal in the Bloch sphere. Using these, a geometric approach can be generally employed. 



All these already give the guessing probability in a simple way for cases when qubit states are given with equal \emph{a priori} probabilities, that is, when $q_{\x}=1/N$ for all $\x$. In this case, it is not difficult to see a general form of the guessing probability. Substituting $q_{\x}=1/N$ in the KKT condition in Eq. (\ref{eq:kkt1}), it is obtained that parameters $\{ r_{\x}\}_{\x=1}^{N}$ are equal. We put $r:=r_{\x}$ for all $\x$. This holds true for arbitrary set of quantum states in general. Then, the guessing probability is written as 
\bea
P_{\g} = \tr[K] = \frac{1}{N} + r,~\mathrm{with}~r = \frac{\| \frac{1}{N} \rho_{\x} - \frac{1}{N} \rho_{\y} \| }{\|  \sigma_{\x} - \sigma_{\y} \| }, ~~~~\label{eq:gpN}
\eea
where the equation for the parameter $r$ is from the relation in Eq. (\ref{eq:kkt1p}), and  the distance measure can be taken as the Hilbert-Schmidt norm $D_{\mathrm{HS}}$ (as it is natural in the Bloch sphere) or the trace norm $D_{T}$. Both measures give the same value of $r$ since they are related only by a constant: $D_{\mathrm{HS}} = \sqrt{2} D_{T}$ for qubit states. This means that, the parameter $r$ can be obtained by either of distance measures while referring to the geometry in the Bloch sphere, since the parameter is only a ratio, see Eq. (\ref{eq:gpN}).

The parameter $r$ corresponds to the ratio between two polytopes, $\P(\{ 1/N, \rho_{\x}\}_{\x=1}^{N})$ of given states and the other $\P(\{\sigma_{\x} \}_{\x=1}^{N})$ only of complementary states, as it is shown in Eq. (\ref{eq:gpN}). We recall that most of $\{\sigma\}_{\x=1}^{N}$ are pure (i.e. rank-one), lying at the border of the Bloch sphere. This implies that the polytope $\P(\{\sigma_{\x} \}_{\x=1}^{N})$ of complementary states is clearly the maximal in the Bloch sphere. In this way, the polytope $\P(\{\sigma_{\x} \}_{\x=1}^{N})$ always contain the origin of the sphere, from which optimal measurements can be constructed. Note also that, from this, the completeness of measurement is fulfilled, see Eq. (\ref{eq:kkt2}). Finally, two polytopes $\P(\{ 1/N, \rho_{\x}\}_{\x=1}^{N})$ and $\P(\{\sigma_{\x} \}_{\x=1}^{N})$ are similar from the relation in Eq. (\ref{eq:kkt1p}) by the ratio $r$, since two polytopes $\P(\{ 1/N, \rho_{\x}\}_{\x=1}^{N})$ and $\P(\{ 1/N, \sigma_{\x}\}_{\x=1}^{N})$ are congruent. 

We summarize the method of finding optimal discrimination in the following.
\begin{description}
\item[ 1] Construct a polytope from given states as the convex hull of $\{1/N,\v(\rho_{\x}) \}_{\x=1}^{N}$ in the Bloch sphere  where vertices correspond to $\v(\rho_{\x})/N$. 
\item[2] Expand the polytope such that it keeps being similar to the original one until it is maximal within the Bloch sphere (as most of $\{\sigma_{\x}\}_{\x=1}^{N}$ are pure), and then compute the ratio $r$ of the resulting polytope with respect to the original one. The guessing probability is thus obtained, $P_{\g} = 1/N +r$.
\item[3] Rotate the maximal polytope within the Bloch sphere until it is found that corresponding lines are parallel to the original ones to fulfill Eq. (\ref{eq:kkt1p}). From this, corresponding vertices are complementary states $\{\sigma_{\x}\}_{\x=1}^{N}$ and optimal POVM elements are explicitly constructed, according to Eq. (\ref{eq:kkt2}).
\end{description}
This completely solves the problem of discrimination of qubit states given with equal \emph{a priori} probabilities.

\begin{figure}[t]
\includegraphics[width=4in]{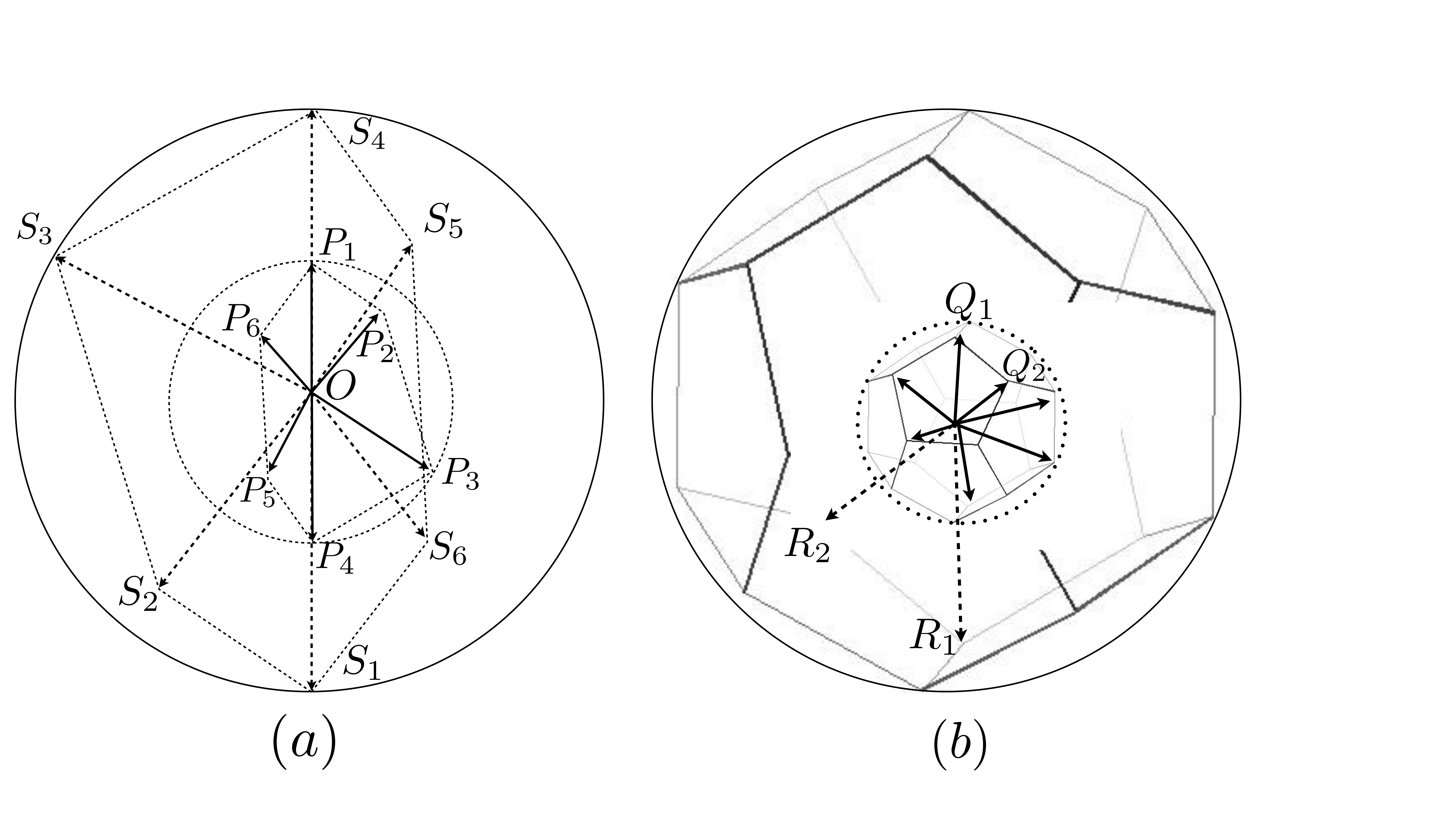}
\caption{ (A) Six states $\{1/6, \rho_{\x}\}_{\x=1}^{6}$ in the half-plane are given with purities $\{f_{\x}\}_{\x=1}^{6}$, i.e., $OP_{\x} = f_{\x} /6$, where three of them ($\x=1,3,4$) have the same purity and the others are less. See also that the relation in Eq. (\ref{eq:kkt1p}) is fulfilled. The ratio $r$ in Eq. (\ref{eq:gpN}) can be obtained by expanding the given polytope until it is maximal in the plane. This is also the ratio between radii of two circles covering respective polygons. Thus, $P_{\g} = 1/6+ f_{1} /6$. Complementary state $\sigma_{\x}$ corresponds to $OS_{\x}$. For $\x=2,5,6$, they are not pure states and thus for these states, the optimal strategy is to make a guess without actual measurement \cite{ref:note1}.  (B) For $\{1/N,\rho_{\x} \}_{\x=1}^{N}$ of pure states, each vertex of the polyhedron corresponds to the Bloch vector of state $\rho_{\x}/N$. The ratio $r$ is equal to $Q_1Q_2 / R_1 R_2$, see also Eq. (\ref{eq:kkt1p}), and thus $P_{\g} =2/N$. Even if these states are modified, if the minimal sphere covering the polyhedron is unchanged, the guessing probability remains the same. }\label{fig1}
\end{figure}

It is already observed that the guessing probability does not depend on detailed relations of given quantum states to discriminate among, but a property from the whole set $\{\rho_{\x} \}_{\x=1}^{N}$, since its ratio $r$ is the relevant parameter. If given states are modified such that the polytope has the same ratio $r$, then the guessing probability remains the same. This means that, in the communication scenario we have introduced in the beginning, Alice who encodes messages can choose, or modify, sets of quantum states $\{\rho_{\x} \}_{\x=1}^{N}$ in such a way that Bob who decodes from quantum states cannot recognize her modification using the optimal guessing. This actually defines equivalence classes of sets of quantum states in terms of optimal guessing \cite{ref:pro}.

In the following, we apply the method to various cases of qubit state discrimination. The simplest example, also the case when a general solution is known, is for $N=2$, say $\rho_1$ and $\rho_2$. Following the instruction in the above, i) the polytope constructed by given two states corresponds to a line connecting two Bloch vectors of the states. The length can be computed using the trace distance as $\|\rho_{1} - \rho_{2} \|/2$. Then, ii) the maximal ploytope similar to the original one is clearly the diameter of the Bloch sphere, which has length $2$ in terms of the trace distance, hence, $r= \|\rho_{1} - \rho_{2} \|/4$ (which equivalently can be obtained with the Hilbert-Schmidt distance). Substituting this in Eq. (\ref{eq:gpN}), the Helstrom bound in Ref, \cite{ref:hel} is reproduced. Then, iii) the diameter can be rotated until it is parallel to the original one. Thus, optimal measurements are also obtained.

Next, let us consider $N$ states on the half-plane. We can begin with $N$ states $\{1/N,\rho_{\x} \}_{\x=1}^{N}$ that are equally distributed in the plane. They are characterized by Bloch vectors: $\v(\rho_{\x}) = f_{\x}(\cos\theta_{\x} , \sin\theta_{\x},0)$ where $\theta_{\x} = 2\pi \x/N$. For these states, no general solution is known, except cases of geometrically uniform states together with the condition that $\{f_{\x}\}_{\x=1}^{N}$ are equal \cite{ref:eldar}. For convenience, we also suppose that $N$ is an even number and assume that $f_{N/2} = f_{N} = \max_{\x} f_{\x}$. Then, applying the method introduced, one can easily find that the ratio depends on the maximal purity, that is, $r=f_{N}/N$, and the guessing probability is obtained as $P_{\g} = 1/N + f_{N}/N$, no matter what purities the other $N-2$ states have, see also Fig. \ref{fig1}. This already reproduces the result in Ref. \cite{ref:eldar} for qubit states. The assumption of the equal distribution on angles can be relaxed while keeping $\theta_{N} = \pi + \theta_{N/2}$ and $f_{N/2} = f_{N} =\max_{\x}f_{\x}$, for which the guessing probability then remains the same no matter how other $N-2$ states are structured. This is because, as it is shown in Eq. (\ref{eq:gpN}), they are given with equal probabilities and the ratio $r$ is unchanged.

Optimal measurements can be analyzed as follows, based on the geometric formulation, see also Fig. \ref{fig1}. For two states $\rho_{N/2}$ and $\rho_{N}$ having the maximal purity, it is clear that measurement is applied, and let $\sigma_{N} = |\psi_{N}\rangle \langle \psi_{N}|$ and $\sigma_{N/2} = |\psi_{N/2}\rangle \langle \psi_{N/2}|$. From these, optimal POVM elements are straightforward. For the other states, say $\{\rho_{\z}\}$ having $f_{\z}<f_{N}$, it holds from the KKT condition in Eq. (\ref{eq:kkt1p}) that $\v_{N} - \v_{\z}$ is parallel to $r(\v(\sigma_{N}) - \v(\sigma_{\z}))$. This simply shows that their complementary states $\{ \sigma_{\z} \}$ cannot be pure states, i.e. not of rank-one. Then, corresponding POVM element is the null operator, that is, for these states the optimal strategy is to make a guess without actual measurement.

The method can be applied to a set of qubit states having a volume. For instance, let us look at the case when pure states are given such that their Bloch vectors form a regular polyhedron of $N$ vertices, see Fig. \ref{fig1}. Following the instruction in the above, the parameter $r$ can be obtained as the ratio of two spheres, one the Bloch sphere and the other the minimal sphere covering the polyhedron of given states $\{1/N,\rho_{\x} \}_{\x=1}^{N}$. From this, we have $r=1/N$, and the guessing probability is thus, $P_{\g}=2/N$. One can also modify angles between those $N$ states such that the minimal sphere covering the polyhedron remains the same, and then the guessing probability is unchanged.

The method of finding optimal state discrimination presented here can be in principle applied to, high dimensional states once their geometry is clear, or qubit states with unequal \emph{a priori} probabilities. For the latter, although the geometry is clear, we do not have yet a general and systematic method to derive the guessing probability. Nevertheless, the geometric formulation can be applied and provide analytic solutions. The guessing probability under equal \emph{a priori} probabilities, which is automatically computed via the geometric formulation, would then give an upper bound. Illuminating examples are presented in Ref. \cite{ref:pro}.

To conclude, we have shown a geometric formulation for qubit state discrimination and provide the guessing probability in a closed form for equal \emph{a priori} probabilities. This makes a significant contribution to the study of quantum state discrimination. Optimal measurements are characterized accordingly. It is shown how qubit states can be modified while the guessing probability remains the same. As qubits are units of quantum information processing, we envisage that the method of discrimination and results presented here would be useful to develop further investigations of qubit applications e.g. Refs. \cite{ref:hel} \cite{ref:ren} \cite{ref:semi}, or approaches to related open questions e.g. Ref. \cite{ref:op}.

This work is supported by National Research Foundation and Ministry of Education, Singapore, and Basic Science Research Program through the National Research Foundation of Korea (NRF) funded by the Ministry of Education, Science and Technology (2010-0007208).



\begin{thebibliography}{99}
\bibitem{ref:note1} The only possibility to choose a positive operator $\sigma_{\x}$ such that $\tr[\sigma_{\x} M_{\x}]=0$ for a positive operator $M_{\x}$ of rank-two is that $\sigma_{\x}=0$. This is taken into account in the the case when $r_{\x}=0$ in the KKT conditions, and therefore excluded. Then, POVM elements are either rank-one or the null operator.
\end{thebibliography}
\end{document}